\documentclass[aps,prd,preprint,superscriptaddress,tightenlines,nofootinbib]{revtex4}

\usepackage{graphicx}
\usepackage{dcolumn}
\usepackage{bm}

\newcommand{\jpsi}{J/\psi}
\newcommand{\pp}{\psi(2S)}
\newcommand{\pizero}{\pi^0}

\newcommand{\ModeA}{\eta_c \to K^+K^-\pi^0}
\newcommand{\ModeB}{\eta_c \to \eta\pi^+\pi^-,\eta\to\gamma\gamma}
\newcommand{\ModeC}{\eta_c \to \eta\pi^+\pi^-,\eta\to\pi^+\pi^-\pi^0}
\newcommand{\ModeD}{\eta_c \to K^+K^-\pi^+\pi^-}
\newcommand{\ModeE}{\eta_c \to \pi^+\pi^-\pi^+\pi^-}
\newcommand{\ModeF}{\eta_c \to K^-\pi^+K^0}

\def\Journal#1&#2&#3(#4){#1{\bf #2}, #3 (#4)}
\def\NIMA{Nucl. Instrum. Methods Phys. Res., Sect. A }
\def\NPB{Nucl.  Phys.  B }
\def\PLB{Phys.  Lett.  B }
\def\PRL{Phys.  Rev.  Lett.  }
\def\PRD{Phys.  Rev.  D }

\def\etal{{\it et al.}}

\begin{document}
\preprint{CLNS 06/1965}       
\preprint{CLEO 06-11}         

\title{Search for $\pp \to \eta_c \pi^+\pi^-\pi^0$}

\author{T.~K.~Pedlar}
\affiliation{Luther College, Decorah, Iowa 52101}
\author{D.~Cronin-Hennessy}
\author{K.~Y.~Gao}
\author{D.~T.~Gong}
\author{J.~Hietala}
\author{Y.~Kubota}
\author{T.~Klein}
\author{B.~W.~Lang}
\author{R.~Poling}
\author{A.~W.~Scott}
\author{A.~Smith}
\author{P.~Zweber}
\affiliation{University of Minnesota, Minneapolis, Minnesota 55455}
% ...
\collaboration{CLEO Collaboration} 
\noaffiliation

\date{November 13, 2006}

\begin{abstract}
Using 5.63 pb$^{-1}$ of data accumulated at the $\pp$ resonance 
with the CLEO III and CLEO-c detectors corresponding to 3.08 
million $\pp$ decays, a search is performed for the decay 
$\pp \to \eta_c \pi^+\pi^-\pi^0$ to test a theoretical prediction 
based upon the assumption that   
the $c\bar{c}$ pair in the $\pp$ does not annihilate directly
into three gluons but rather survives before annihilating.
No signal is observed, and a combined upper limit from six 
$\eta_c$ decay modes is determined 
to be ${\cal B} (\pp \to \eta_c \pi^+\pi^-\pi^0) \leq 1.0\times10^{-3}$ 
at 90\% C.L. This upper limit is about an order of magnitude 
below the theoretical expectation. 

\end{abstract}

\maketitle
In perturbative QCD the charmonium states $\jpsi$ and $\pp$ are
nonrelativistic bound states of a charm and an anticharm quark 
and it is predicted that 
the decays of these states are dominated by the 
annihilation of the charm and anticharm quark into three gluons. 
The partial width for the decays into an exclusive hadronic state $h$
is then expected to be proportional to the square of the
$c\bar{c}$ wave function overlap at zero quark separation,
which is well determined from the leptonic width~\cite{PDG}. 

Since the strong coupling constant, $\alpha_s$, is not very
different at the $J/\psi$ and $\psi(2S)$ masses, it is expected that
for any state $h$ the $J/\psi$ and $\psi(2S)$ branching fractions are 
related by~\cite{RULE}
\footnotesize
\begin{equation}
Q_h=\frac{{\cal B}(\psi(2S)\to h)}{{\cal B}(J/\psi\to h)}
\approx
\frac{{\cal B}(\psi(2S)\to \ell^+\ell^-)}{{\cal B}(J/\psi\to\ell^+\ell^-)}
=(12.4 \pm 0.4)\%,
\label{equ:q}
\end{equation}
\normalsize
where ${\cal B}$ denotes a branching fraction,
and the leptonic branching fractions are taken from the 
Particle Data Group (PDG)~\cite{PDG}.
This relation is sometimes called \lq \lq the 12\% rule''. Modest
deviations from the rule are expected~\cite{GULI}. Although the
rule works well for some specific decay modes, 
isospin conserving $\psi(2S)$ decays to two-body final states 
consisting of one vector and one pseudsoscalar meson exhibit strong
suppression: $\rho\pi$ is suppressed by a factor of seventy compared 
to the expectations of the rule (the so-called $\rho\pi$ puzzle)
~\cite{PDG,MOREBES,HARRIS,BHHMK}.
Also, vector-tensor channels such as $\rho a_2(1320)$, 
and $K^*(892)\bar{K}^*_2(1430)$ are significantly
suppressed \cite{PDG,BESVT}.
Another issue is the hadronic excess in $\pp$ decays:
the inclusive hadronic decay rate of $\pp$  is larger than that
expected from an extrapolation of the $J/\psi$ hadronic decay 
branching fraction by $60-70\%$.
A recent review~\cite{GULI} concludes that current theoretical 
explanations of $\pp$ decays are unsatisfactory and that more 
experimental measurements are desirable.

A recent paper \cite{survival} suggests that 
"survival before annihilation" could be an important mechanism 
in $\pp$ decays. This model proposes that
the $c\bar{c}$ pair in the $\pp$ does not annihilate directly
into three gluons but rather "survives" before annihilating,
{\it i.e}., the $c\bar{c}$ pair decays by first emitting two or three
nonperturbative gluons before annihilating
into three or two perturbative gluons.
This model, it is claimed, can solve the problem of the 
apparent hadronic excess in $\pp$ decays as well as the 
$\rho\pi$ puzzle. One important prediction of the model is that 
the $\pp \to \eta_c \pi^+\pi^-\pi^0$ channel would be a significant 
decay with a branching fraction of 1\% or larger.

We search for the decay 
$\pp \to \eta_c \pi^+\pi^-\pi^0$ using the six decay
modes of the $\eta_c$ listed in Table \ref{tab:num}.
The $\eta_c$ decay modes selected amount to about 
10.6\% of the total $\eta_c$ decay rate \cite{PDG}.

The data sample used in this analysis was obtained at the $\psi(2S)$ in 
$e^+e^-$ collisions produced by the Cornell Electron Storage Ring (CESR) 
and acquired with the CLEO detector.
The CLEO~III detector~\cite{cleoiiidetector} features a solid angle
coverage for charged and neutral particles of 93\%.
The charged particle tracking system, operating in a
1.0~T magnetic field along the beam axis, achieves
a momentum resolution of $\sim$0.6\% at
$p=1$~GeV/$c$. The calorimeter attains a photon
energy resolution of 2.2\% at $E_\gamma=1$~GeV and 5\% at 100~MeV.
Two particle identification systems, one based on energy loss ($dE/dx$) 
in the drift chamber and the other a ring-imaging Cherenkov (RICH)
detector, are used together to separate $K^\pm$ from $\pi^\pm$.
The combined $dE/dx$-RICH particle identification procedure has
efficiencies exceeding 90\% and misidentification rates below 5\%
for both $\pi^\pm$ and $K^\pm$ for momenta below 2.5 GeV/$c$.

Half of the $\psi(2S)$ data were taken after a transition 
to the CLEO-c~\cite{YELLOWBOOK} detector configuration, 
in which the CLEO silicon small radius tracking detector 
was replaced with a six-layer all-stereo drift chamber.
The two detector configurations also correspond
to different accelerator lattices: the former with
a single wiggler magnet and a center-of-mass
energy spread of 1.5~MeV, the latter
(CESR-c~\cite{YELLOWBOOK}) with
six wiggler magnets and an energy spread of 2.3~MeV.

The integrated luminosity ($\cal{L}$) of the datasets was measured
using $e^+ e^-$, $\gamma\gamma$, and $\mu^+ \mu^-$ final
states~\cite{LUMINS}. Event counts were normalized with a Monte
Carlo (MC) simulation based on the Babayaga~\cite{BBY} event
generator combined with GEANT-based~\cite{GEANT} detector
modeling. The data consist of  $\cal{L}$=5.63~pb$^{-1}$ on the
peak of the $\psi(2S)$ at $\sqrt s = 3.686$ GeV 
(2.74~pb$^{-1}$ for CLEO~III, 2.89~pb$^{-1}$ for CLEO-c). 

Standard requirements \cite{multibody} are used to select charged 
particles reconstructed in the tracking system and photon candidates 
in the CsI calorimeter. We require tracks of charged particles
to have momenta $p>100$~MeV/$c$ and to satisfy $|\cos\theta|<0.90$, 
where $\theta$ is the polar angle with respect to the $e^+$ direction.
Each photon candidate satisfies  $E_\gamma>30$~MeV and is more than 
8\,cm away from the projections of tracks into the calorimeter.
Particle identification from $dE/dx$ and the RICH detector is used 
for all charged particle candidates. Pions and kaons must be 
positively and uniquely identified, i.e.,
pion candidates must not satisfy kaon or proton selection criteria,
and kaon candidates obey similar requirements.

The invariant mass of the decay products from the following particles
must lie within limits determined from MC studies:
$\pi^0~(120 \le M_{\gamma\gamma} \le 150 {\rm ~MeV})$, 
$\eta ~(500 \le M_{\gamma\gamma} \le 580 {\rm ~MeV})$, 
$\eta ~(530 \le M_{\pi^+\pi^-\pi^0} \le 565 {\rm ~MeV})$.

For $\pi^0 \rightarrow \gamma \gamma$
and $\eta \rightarrow \gamma \gamma$ candidates in events with
more than two photons, the combination giving a mass closest to
the known  $\pi^0$ or $\eta$ mass is chosen, and a kinematically
constrained fit to the known parent mass is made. 
Fake $\pi^0$ and $\eta$ mesons are suppressed by 
requiring that each electromagnetic shower profile be 
consistent with that of a photon.
For $\eta \rightarrow \pi^+ \pi^- \pi^0$ 
the $\pi^0$ is selected as described above, and then
combined with all possible combinations of two oppositely charged
pions choosing the combination that is closest to the $\eta $ mass.
A kinematically constrained fit is not used for this mode.

Events having final state particles consistent with one of the 
six $\eta_c$ decay channels and additionally a $\pi^+\pi^-\pi^0$ 
combination are selected. Energy and momentum conservation 
requirements are then imposed on the event using the summed 
vector momentum $\sum {\bf p_i}$ and scalar energy $E_{\rm vis}$. 
These requirements are based on the experimental resolutions 
as determined by Monte Carlo for each of the final states. 
The scaled momentum $|\sum {\bf p_i}|/E_{\rm c.m.}$ is required to be
consistent with zero and the scaled energy $E_{\rm vis}$/$E_{\rm c.m.}$
is required to be consistent with unity. The experimental resolutions 
are less than 1\% in scaled energy and 2\% in scaled momentum.
In order to veto the final states 
$\pp \to X J/\psi$ ($X= \pi^+ \pi^-$, $\pi^0 \pi^0$, or $\eta$)
events are rejected in which the mass of any of the   
following falls within the range $3.05<m<3.15$ GeV:
the two highest momentum oppositely charged tracks,
the recoil mass against the two lowest momentum oppositely charged tracks,
or the mass recoiling against the $\pi^0\pi^0$ or $\eta$.

The efficiency, $\varepsilon$, for each final state is the weighted average
obtained from MC simulations~\cite{GEANT} for both detector configurations.
No initial state radiation is included in the Monte Carlo, but final state
radiation is accounted for. The efficiencies in Table \ref{tab:num} include
the branching fractions of the $\eta$ \cite{PDG}.

\begin{figure}[htbp]
  \centering
  \includegraphics[height=0.32\textheight]
   {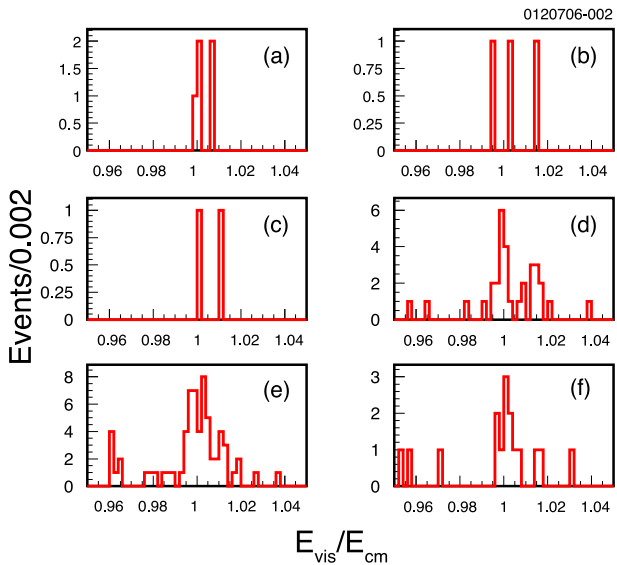}
  \caption{The scaled total energy distribution for each $\pp$ mode
 for events with the candidate $\eta_c$ decay having an invariant mass 
 greater than 2.7 GeV.
 a) $\eta_c \to K^+K^-\pi^0$.
 b) $\eta_c \to \eta\pi^+\pi^-, \eta\to\gamma\gamma$.
 c) $\eta_c \to \eta\pi^+\pi^-, \eta\to\pi^+\pi^-\pi^0$.
 d) $\eta_c \to K^+K^-\pi^+\pi^-$.
 e) $\eta_c \to \pi^+\pi^-\pi^+\pi^-$.
 f) $\eta_c \to K^-\pi^+K^0$.}
\label{xtot}
\end{figure}

\begin{figure}[htbp]
  \centering
  \includegraphics[height=0.32\textheight]
   {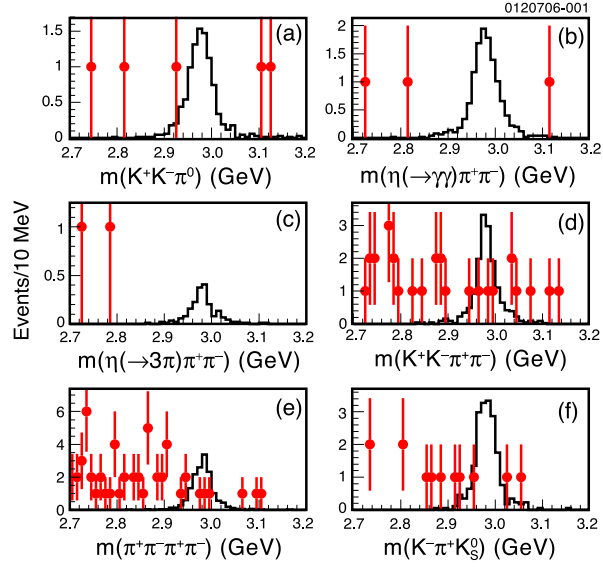}
  \caption{The invariant mass distribution for each $\eta_c$ mode.
 Histogram: Monte Carlo ($y$ axis on left), normalized to 
 ${\cal B} (\pp \to \eta_c \pi^+\pi^-\pi^0)$ = 1\%;
 filled circles with error bar: data ($y$ axis on right).
 a) $\eta_c \to K^+K^-\pi^0$.
 b) $\eta_c \to \eta\pi^+\pi^-, \eta\to\gamma\gamma$.
 c) $\eta_c \to \eta\pi^+\pi^-, \eta\to\pi^+\pi^-\pi^0$.
 d) $\eta_c \to K^+K^-\pi^+\pi^-$.
 e) $\eta_c \to \pi^+\pi^-\pi^+\pi^-$.
 f) $\eta_c \to K^-\pi^+K^0$.}
\label{mass}
\end{figure}

\begin{figure}[htbp]
  \centering
  \includegraphics[height=0.35\textheight]
   {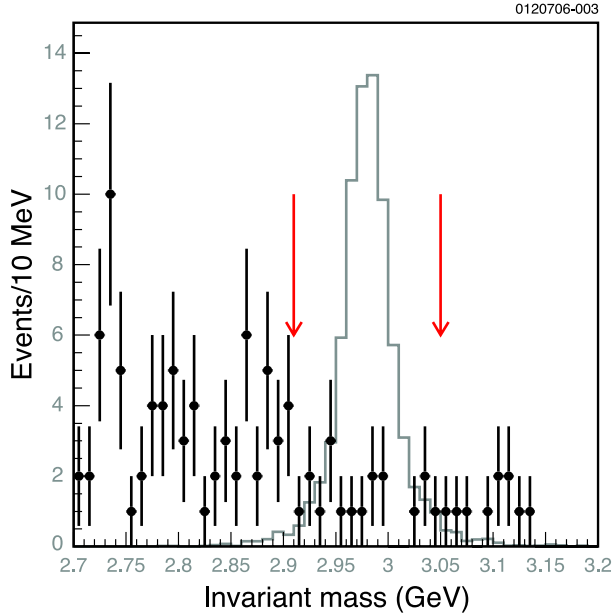}
  \caption{The invariant mass distribution for all six $\eta_c$ modes combined. 
           Histogram: Monte Carlo, normalized to 
           ${\cal B} (\pp \to \eta_c \pi^+\pi^-\pi^0)$ = 1\%;
           filled circles with error bar: data.}
  \label{combined6modes}
\end{figure}

After the selection of events consistent with the six exclusive $\pp$ 
decay modes, we search for $\eta_c$ production in these events by 
examining the invariant mass of combinations of particles consistent 
with an $\eta_c$ decay.
In all events there are multiple $\eta_c$ combinations and to be
conservative we have taken only one combination per event choosing the one
nearest the $\eta_c$ mass. Figure \ref{xtot} is the scaled energy 
distribution for each of the six $\pp$ exclusive decays after imposing 
momentum conservation ($|\sum {\bf p_i}|/E_{\rm c.m.} < 0.02$)
and for events with an $\eta_c$ candidate mass 
greater than 2.7 GeV, which includes the nominal $\eta_c$ mass region. 
There is clear evidence of exclusive $\pp$ production of the final states 
under study from the accumulation of events near unity. 
After requiring the scaled energy to be in the range $(0.98-1.02)$, 
the invariant mass distributions shown in Fig. \ref{mass} are 
analyzed for $\eta_c$ production. A combined mass distribution for all 
six modes is shown in Fig. \ref{combined6modes}. The histograms show 
the expected signal from Monte Carlo normalized to the branching ratio 
prediction of  ${\cal B} (\pp \to \eta_c \pi^+\pi^-\pi^0 = 1\%$ . 
There is no evidence for $\eta_c$ production in any of the six individual 
decay modes or in the combined distribution.

Table \ref{tab:num} shows the number of events in the $\eta_c$ mass
region ($2.91 - 3.05$ GeV), which corresponds to $\pm 3$ standard 
deviations of the expected $\eta_c$ mass resolution, the efficiency 
and the $\eta_c$ branching ratio for each of the decay channels.
No evidence is found for $\pp \to \eta_c \pi^+\pi^-\pi^0$.

\begin{table}[htbp]
\caption{No evidence is found for $\pp \to \eta_c \pi^+\pi^-\pi^0$.
         Number of events in the $\eta_c$ mass region, detection efficiency,
         and $\eta_c$ branching ratio~\cite{PDG}.}
\begin{center}
\begin{tabular}{|l|c|c|c|}  \hline
decay mode & \#events & efficiency (\%) & ${\cal B}(\eta_c)$ (\%) \\ \hline
 $\ModeA$  &  1 & 3.08 & $1.2\pm0.2$   \\
 $\ModeB$  &  0 & 2.76 & $1.3\pm0.5$ \\
 $\ModeC$  &  0 & 0.80 & $0.7\pm0.3$ \\
 $\ModeD$  &  7 & 3.07 & $1.5\pm0.6$     \\
 $\ModeE$  &  6 & 4.06 & $1.2\pm0.3$     \\
 $\ModeF$  &  4 & 1.55 & $4.7\pm0.8$     \\ \hline
all modes combined
           & 18 & 2.31 &$10.6\pm1.2$  \\ \hline
\end{tabular}
\label{tab:num}
\end{center}
\end{table}

The following sources of systematic uncertainties have been considered.
The number of $\psi(2S)$ decays (3.0\%), trigger efficiency (1.0\%),
and the uncertainty associated with the resolution defining
a signal region in the scaled energy and resonant mass (2.0\%)
contribute identical systematic uncertainties to each channel.
Other sources vary by channel, for example, 
Monte Carlo statistics (2.7\% $-$ 7.4\%).
The systematic uncertainty associated with the charged track 
finding is 0.7\% per track. 
This uncertainty adds coherently for each charged track in the event. 
The particle identification uncertainty is 0.3\% per charged pion, 
and 1.3\% per charged kaon. 
This uncertainty adds coherently for each charged track in the event 
since PID is used for every track.
The systematic uncertainty associated with
$\pizero$ and $\eta\to\gamma\gamma$ finding is  4.4\%. 

We determine a $\pp \to \eta_c \pi^+\pi^-\pi^0$ branching fraction 
upper limit at 90\% C.L. combining all $\eta_c$ decay modes using
\begin{equation}
{\cal B}=\frac{N_{S}}
{(\sum\varepsilon_i \cdot {\cal B}(\eta_c\to h_i)) \cdot N_{\pp}},
\end{equation}
where $N_{S}$ is the upper limit on the number of signal events,
$\varepsilon_i$ and ${\cal B}(\eta_c\to h_i)$ 
are the efficiency and the branching fraction for $\eta_c$ decay mode $i$;
and $N_{\pp}$ is the total number of $\pp$ decays.
The number of $\pp$ decays $N_{\pp}$ was determined to be
$3.08\times 10^6$ by the method described in \cite{cbx04-35}.
The combined efficiency, which includes the branching fractions for $\eta_c$ 
decays, is defined as
\begin{equation}
\varepsilon=\frac{\sum\varepsilon_i \cdot {\cal B}(\eta_c\to h_i)}
                 {\sum{\cal B}(\eta_c\to h_i)},
\end{equation}
and is also shown in Table~\ref{tab:num}.
To determine the upper limit on the number of signal events, $N_S$, 
the data distribution in Fig.~\ref{combined6modes} is fit using a 
polynomial and a Gaussian with fixed signal shape obtained from MC. 
By fixing the signal amplitude and maximizing the fit  
likelihood with respect to the parameters for background,
we obtain the fit likelihood as a function of the signal
amplitude and determine the upper limit on the number of 
signal events to be $< 6.7$ at 90\% C.L.
After combining the systematic uncertainties and the 
uncertainties in the $\eta_c$ branching ratios in 
quadrature and increasing the upper limit on the number 
of observed events by one standard deviation of the combined 
uncertainty, the branching fraction upper limit is 
computed to be $1.0\times10^{-3}$.
[Alternative methods of computing a 90\% C.L. upper limit 
using the number of events in the signal and sideband regions 
of the invariant mass give slightly lower limits \cite{Feldman, PDG}.]


In summary we have searched for the decay $\pp \to \eta_c \pi^+\pi^-\pi^0$ 
using six decay modes of the $\eta_c$. We have determined a combined 
upper limit at 90\% confidence level for the branching fraction of the 
decay $\pp \to \eta_c \pi^+\pi^-\pi^0$ to be $1.0\times10^{-3}$. 
This upper limit is about an order of 
magnitude below the theoretical prediction of the survival before 
annihilation model meaning that the survival of the $c\bar{c}$ pair 
to form an $\eta_c$ is highly suppressed.

We gratefully acknowledge the effort of the CESR staff 
in providing us with excellent luminosity and running conditions. 
D.~Cronin-Hennessy and A.~Ryd thank the A.P.~Sloan Foundation. 
This work was supported by the National Science Foundation,
the U.S. Department of Energy, and 
the Natural Sciences and Engineering Research Council of Canada.


\begin{thebibliography}{99}

\bibitem{PDG} Particle Data Group, 
W.-M. Yao {\it et al.}, J. Phys. G {\bf 33}, 1 (2006)

\bibitem{RULE}W. S. Hou and A. Soni, {\Journal\PRL&50&569(1983)};
        G. Karl and W. Roberts, {\Journal\PLB&144&263(1984)};
        S. J. Brodsky \etal, {\Journal\PRL&59&621(1987)};
        M. Chaichian \etal, {\Journal\NPB&323&75(1989)};
        S. S. Pinsky, {\Journal\PLB&236&479(1990)};
        X. Q. Li \etal, {\Journal\PRD&55&1421(1997)};
        S. J. Brodsky and M. Karliner, {\Journal\PRL&78&4682(1997)};
        Yu-Qi Chen and Eric Braaten, {\Journal\PRL&80&5060(1998)};
        M. Suzuki, {\Journal\PRD&63&054021(2001)};
        J. L. Rosner, {\Journal\PRD&64&094002(2001)};
        J. P. Ma, {\Journal\PRD&65&097506(2002)};
        M. Suzuki, {\Journal\PRD&65&097507(2002)}.

\bibitem{GULI} Y. F.~Gu and X. H.~Li, {\Journal\PRD&63&114019(2001)}.

\bibitem{BHHMK} CLEO Collaboration, 
N. E. Adam {\it et al.}, {\Journal\PRL&94&012005(2005)}.

\bibitem{MOREBES}  BES Collaboration,
M. Ablikim {\it et al.},
{\Journal\PLB&614&37(2005)}, 
{\Journal\PLB&598&149(2004)}, 
{\Journal\PRL&93&112002(2004)}, 
{\Journal\PRD&72&012002(2005)}; 
{\Journal\PLB&619&247(2005)}, 
and
{\Journal\PRD&70&012003(2004)}; 
BES Collaboration, J. Z.~Bai {\it et al.},
{\Journal\PRD&70&012006(2004)}. 

\bibitem{HARRIS} F. A. Harris, 
Int. J. Mod. Phys. A {\bf 20}, 445 (2005).

\bibitem{BESVT}  BES Collaboration, J. Z.~Bai {\it et al.}, 
{\Journal\PRD&69&072001(2004)}.

\bibitem{survival} P. Artoisenet \etal, {\Journal\PLB&628&211(2005)}.
 
\bibitem{cleoiiidetector} CLEO Collaboration, Y. Kubota {\it et al.}, 
{\Journal\NIMA&320&66(1992)};
D. Peterson {\it et al.}, {\Journal\NIMA&478&142(2002)};
M.~Artuso {\it et al.}, {\Journal\NIMA&554&147(2005)}. 

\bibitem{YELLOWBOOK} CLEO-c/CESR-c Taskforces \& CLEO-c Collaboration, 
Cornell University LEPP Report No. CLNS~01/1742, 2001 (unpublished).

\bibitem{LUMINS} CLEO Collaboration, G.~Crawford {\it et al.},
{\Journal\NIMA&345&429(1994)}.

\bibitem{BBY} C.~M.~Carloni Calame {\it et al.},
  Nucl. Phys. B, Proc. Suppl. {\bf 131}, 48 (2004).

\bibitem{GEANT} Computer code GEANT~3.21, in R.~Brun {\it et al.}, 
CERN Report No. W5013, (1993) (unpublished).

\bibitem{multibody} CLEO Collaboration, R. A. Briere  {\it et al.},
{\Journal\PRL&95&062001(2005)}.

\bibitem{cbx04-35} CLEO Collaboration, S. B. Athar {\it et al.},
{\Journal\PRD&70&112002(2004)}.

\bibitem{Feldman} G. J. Feldman and R. D. Cousins,
{\Journal\PRD&57&3873(1998)}.

\end{thebibliography}
\end{document}